\documentclass[structabstract]{aa} 
\usepackage{graphicx}
\usepackage{amssymb}
\usepackage{subfigure}
\usepackage{txfonts}
\usepackage{natbib}
\usepackage{lscape} % stellenweises Querformat
\begin{document}

   \title{Coronal properties of planet-bearing stars}

   \author{K.~Poppenhaeger
      \and
       J.~Robrade
      \and
          J.~H.~M.~M.~Schmitt}
   \institute{Hamburger Sternwarte, University Hamburg,
             Gojenbergsweg 112, 21029 Hamburg, Germany\\
              \email{katja.poppenhaeger@hs.uni-hamburg.de}}

   \date{Received 11 February 2010; accepted 20 March 2010}

  \abstract
{Do extrasolar planets affect the activity of their host stars? Indications for chromospheric activity enhancement have been found for a handful of targets, but in the X-ray regime, conclusive observational evidence is still missing.}{We want to establish a sound observational basis to confirm or reject major effects of Star-Planet Interactions (SPI) in stellar X-ray emissions.}{We therefore conduct a statistical analysis of stellar X-ray activity of all known planet-bearing stars within 30~pc distance for dependencies on planetary parameters such as mass and semimajor axis.}{In our sample, there are no significant correlations of X-ray luminosity or the activity indicator $L_X/L_{bol}$ with planetary parameters which cannot be explained by selection effects.}{Coronal SPI seems to be a phenomenon which might only manifest itself as a strong effect for a few individual targets, but not to have a major effect on planet-bearing stars in general.}


\maketitle

% \abstract{}{}{}{}{} % 5 {} token are mandatory

%_______________________________________________________________

\section{Introduction}

%Allgemeines, SPI: Kashyap, Shkolnik und so weiter. Warum interessant? -> Evaporation, habitable zones, magnetic fields of stars and planets.

The detection of extrasolar planets is one of the outstanding achievements in astronomy during the last 20 years. The first detected exoplanet revealed properties which were very surprising at that time: {\em 51~Peg} \citep{mayorqueloz1995} hosts a Jupiter-like planet at a distance of only 0.05~AU, thus the planet orbits its host star in less than five days. Since then, more than 400~other exoplanets have been found at the time of writing (see for example the Exoplanet Database at www.exoplanet.eu), both in very close orbits and in such more familiar from our own Solar system. 

With the existence of extrasolar planets established, the question arises what the environmental properties of such planets may be and if they might even allow the existence of life. The physical properties of planets, especially in close orbits, are crucially determined by the irradiation from their host stars. The evaporation rate of a planetary atmosphere depends on its exospheric temperature $T_{\infty}$ , i.e., the regions where particles can escape freely \citep{Lammer2003}. Thus, the host star's EUV and X-ray radiation is the key property determining a planet’s exospheric temperature. Evaporation of the planetary atmosphere has been observed for the transiting planet {\em HD~209458b} \citep{Vidal-Madjar2003}: the planet loses hydrogen which is observable in absorption spectra during the transit.

At very close distances, one might expect also planets to influence their host stars, in analogy to binary stars which show a higher activity level compared to single stars. Two different processes for Star-Planet-Interaction (SPI) have been put forward \citep{CuntzSaar2000}. Planets can induce tidal bulges on the star with an interaction strength depending on the planetary semimajor axis ($\propto a_{pl}^{-3}$), which might lead to enhanced coronal activity via increased turbulence in the photosphere. Planetary magnetic fields can also interact with the stellar magnetic field ($\propto a_{pl}^{-2}$) and might also induce enhanced activity via Jupiter-Io-like interaction, i.e. flux tubes which connect star and planet and heat up their footpoints on the stellar surface, or magnetic reconnection. Some observational campaigns have been conducted to investigate the existence of possible SPI: \cite{Shkolnik2005} monitored the chromospheric activity of 13 stars via \ion{Ca}{ii} H and K line fluxes and found indications for cyclic activity enhancements in phase with the planetary orbit for two of these stars. The activity enhancements appeared once per planetary orbit, suggesting magnetic instead of tidal interaction. However, measurements obtained three years later \citep{Shkolnik2008} showed that the activity enhancements had switched to a cycle in phase with the stellar rotation period instead.

The {\em coronal} activity of planet-bearing stars has been investigated in a first systematic study by \cite{kashyapdrakesaar2008}. The authors claim an over-activity of planet-bearing stars of a factor of four compared to stars without planets, but their study had to include upper limits for a large number of stars since less than one third of the stars in their original sample were detected in X-rays at that time. A dedicated campaign to search for magnetic SPI in the case of {\em HD~179949}, one of the stars which \cite{Shkolnik2005} found to have cyclic activity changes in the chromosphere, was conducted by \cite{SaarSPI2008}. These authors found spectral and temporal variability phased with the planetary orbit, but some of that might also be induced by intrinsic stellar activity variations, since the stellar rotation period is poorly known ($P_*=7-10$~d).

Up to now, the observational basis of stellar coronal activity enhancements due to close-in planets is not sound enough to establish or reject the possibility of coronal SPI. In order to adress this issue we conducted an X-ray study of all planet-hosting stars within a distance of 30~pc with XMM-Newton which have not been studied with ROSAT before. In this fashion a volume-limited complete stellar sample can be constructed.

\section{Observations and data analysis}\label{analysis}

As of December 1st 2009, a total of 72 stars within 30~pc distance have been detected which are known to to harbor one or more planets. For some of these, X-ray properties are known from previous ROSAT or XMM-Newton observations, but for a large number of these stars X-ray characteristics were not or only poorly known. Therefore we observed a total of 20~planet-hosting stars with XMM-Newton between May~2008 and April~2009 to determine X-ray luminosities for stars which had not been detected before in other X-ray missions, and to derive coronal properties from spectra recorded with EPIC (both MOS and PN CCD detectors) especially for stars with close-in planets. We reduced the data with SAS version 8.0, using standard criteria for filtering the data. We extracted counts from the expected source regions with radii between 10'' and 30'', depending on the source signal, background conditions and the presence of other nearby sources. Background counts were extracted from much larger, source-free areas on the same chip for the MOS detectors and at comparable distances from the horizontal chip axis for the PN detector.

For hitherto undetected stars, which showed only a weak source signal in our observations, we used the source detection package "edetect-chain" of SAS v8.0. As stars with low X-ray luminosities in general have lower coronal temperatures and thus softer spectra, we used energy bands of 0.2-1~keV in PN and 0.15-1~keV in MOS and merged all EPIC detectors for source detection.

For the subsequent analysis of all stars we use the four energy bands 0.2-0.45~keV, 0.45-0.75~keV, 0.75-2~keV and 2-5~keV, because not all of our sample stars were detected with sufficient signal to noise ratio to allow spectral fitting. With the four energy bands, we can calculate the stellar fluxes via ECFs (energy conversion factors) for each band more accurately than by just assuming a single ECF for all counts. Above 5~keV, there is very little to no signal present in comparison to softer energies for all of our stars. We calculated these ECFs by simulating spectra in Xspec~v12.5 for different coronal temperatures with the respective instrumental responses and effective areas of the detectors folded in. This yields reliable ECFs which vary about 25\% for coronal temperatures above 1~MK for thin and medium filters. For the thick filter, the small effective area below 350~eV introduces larger errors in the ECFs already for temperatures below $\log T [K] = 6.2$.

For the error estimate on our derived luminosities, we use Poissonian errors on the total number of source counts, and an additional error of 30\% to account for uncertainties in the ECFs and stellar variability. For stars which were not observed with XMM-Newton, we use the published X-ray luminosities from \cite{kashyapdrakesaar2008} and add an extra error of 40\% on top of their Poissonian errors, since \cite{kashyapdrakesaar2008} used a single ECF for their flux calculations.

When comparing X-ray luminosities derived from XMM-Newton and ROSAT observations, one has to take into account the different energy bands accessible to the detectors (0.2-12~keV for XMM-Newton, 0.1-2.4~keV for ROSAT). For coronal temperatures between $\log T = 6.2$ and $7.0$, the flux in the ROSAT band is larger by a factor of 1.1 compared to the XMM-Newton band. For lower coronal temperatures between $\log T = 6.0$-$6.2$ and $\log T = 5.8$-$6.0$, the flux in the ROSAT band is larger by a factor of $\approx 1.5$ and  $\approx 4.0$, respectively, rising steeply towards even lower temperatures, since the spectrum shifts to energies which are inaccessible to  XMM-Newton. With these factors, we can transform the XMM-Newton fluxes to the ROSAT band, identifying stars with coronal temperatures below $\log T = 6.2$ and $6.0$ by a hardness ratio of $HR = (H-S)/(H+S)< +0.19$ and $-0.34$ respectively, with $H$ and $S$ being the source counts in the energy bands $450-750$~eV and $200-450$~eV, respectively. 
% Kashyap data: energy band 0.1-4.5 keV

We did not exclude flaring periods of individual stars when doing comparisons of X-ray luminosities or activity indicators, since we cannot identify flares in stars which are barely detectable and do not allow lightcurve analysis. We do, however, distinguish between flaring and quiescent phases for spectral analyses of individual stars.

When conducting intra-sample comparisons, we will use only detections for Kolmogorov-Smirnov and correlation tests, but we will include upper limits when doing linear regressions of X-ray luminosities or activity indicators over planetary properties.

\section{Sample properties}

Now we characterize our sample of planet-bearing stars within 30~pc with respect to X-ray detection rates and X-ray surface fluxes of ROSAT- and XMM-Newton-detected stars as well as in comparison with field stars. Tables~\ref{counts} and \ref{rosatcounts} list stellar and planetary parameters as well as X-ray properties of the sample stars which have been detected with XMM-Newton and ROSAT, respectively.

\subsection{X-ray detection rate}

In total, 72 stars planet-bearing stars have been detected within a distance of 30~pc. 36 of these were observed with XMM-Newton over the last years, yielding 32 X-ray detections. For 24 additional stars, which have not been observed with XMM-Newton, X-ray luminosities are known from ROSAT observations. This yields 56 stars with known $L_X$ out of the total sample of 72 stars. In our further sample analysis, we will leave three detected stars out of our analysis, namely $\gamma$~Cep, Fomalhaut and $\beta$~Pic; the former being a spectroscopic binary which cannot be resolved in X-rays, the two latter being A-type stars for which the production process of X-ray emission is supposedly very different from later-type stars with a corona and therefore also any planetary influence on X-ray properties should be determined by a different mechanism compared to stars of spectral type F and later.

\begin{figure}
\includegraphics[width=0.5\textwidth]{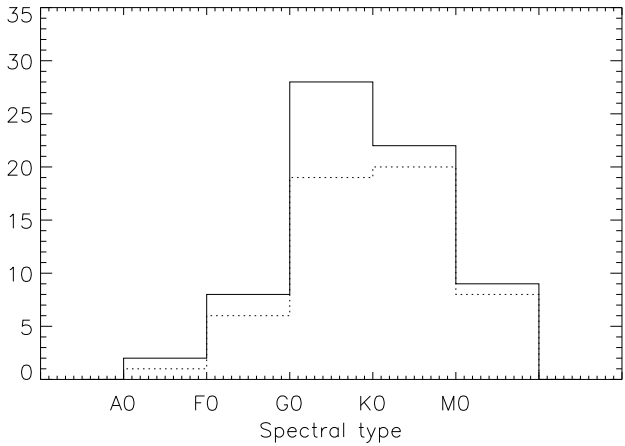}
\caption{Spectral types of planet-bearing stars within 30~pc (solid); X-ray detections marked as dotted lines.}
\label{histo_spectype}
\end{figure}

The stars within 30~pc around which planets have been detected are mainly of spectral type G or later. Fig.~\ref{histo_spectype} gives the rate of X-ray detections versus spectral type, being $75\%$ for F stars, $>65\%$ for G stars and $>85\%$ for K and M stars.

%%%%%%%%%%%%%%%%%%%%%%%%%%%%%%%%%%%
%%%%%%%%%%%%%%%%%%%%%%%%%%%%%%%%%%%

%%%%%%%%%%%%%%%%%%%%%%%%%%%%%%%%%%%
\begin{figure}
\includegraphics[width=0.5\textwidth]{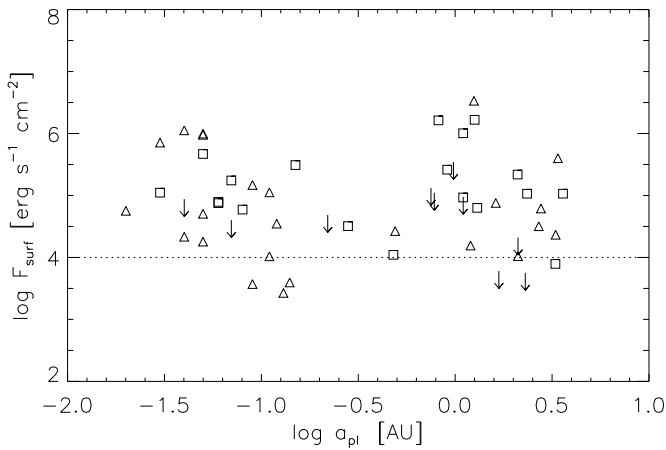}
\caption{X-ray surface flux of planet-bearing stars vs.~planetary distance. XMM-Newton fluxes are shown as triangles, ROSAT fluxes as squares. XMM-Newton fluxes scaled to the ROSAT energy band; the flux level of a solar coronal hole ($\log F_{surf} \approx 4$) is indicated by the dotted line.}
\label{Fx}
\end{figure}
%%%%%%%%%%%%%%%%%%%%%%%%%%%%%%%%%%%

\subsection{X-ray surface flux}

For a subsample of our stars excluding giants, we examined the X-ray surface flux. The lowest flux level of XMM-Newton-detected stars seems to be systematically lower than for ROSAT-detected stars. This is not surprising, since both X-ray telescopes have different accessible energy bands (0.2-12.0~keV for XMM-Newton, 0.1-2.4~keV for ROSAT) and the integrated X-ray flux depends strongly on the lower energy cutoff especially for cool coronae as in our stars. 

Gauged to the same energy band (as described in section~\ref{analysis}), both XMM-Newton- and ROSAT-detected stars show a limiting X-ray surface flux level near $\log F_{surf}\, [{\rm erg\, s^{-1}\, cm^{-2}}] \gtrsim 4.0$ (see Fig.~\ref{Fx}). For the calculation of $F_{surf}$, we use the stellar radii given in the exoplanet.eu database. If we compare the XMM-Newton and ROSAT surface flux sample with the Kolmogorov-Smirnov test, we find that both populations are significantly different. This is due to the selection effect that we proposed planet-bearing stars which were previously undetected with ROSAT (and therefore have low X-ray luminosities) for detection pointings with XMM-Newton. This leads to a higher concentration of stars near the limiting surface flux level of $\log F_{surf}\, [{\rm erg\, s^{-1}\, cm^{-2}}] \approx 4$. For the XMM-Newton and ROSAT subsamples of stars with a surface flux above $\log F_{surf}\, [{\rm erg\, s^{-1}\, cm^{-2}}] \geq 4.5$, we find that these populations are statistically indistinguishable (probability for both samples stemming from the same distribution $71\%$).

%%%%%%%%%%%%%%%%%%%%%%%%%%%%%%%%%%%
\begin{table}
\begin{tabular}{l l c c}
    \hline \hline
Parameters & Data set & Spearman's $\rho$ & Probability $p$\\
    \hline
$L_X$ with $a_{pl}$			& X 		& -0.05	& 0.81 \\
					& X + R		& -0.02 & 0.91 \\
$L_X/L_{bol}$ with $a_{pl}$		& X 		& -0.12 & 0.54 \\
					& X + R		& -0.11 & 0.43 \\
$L_X$ with $M_{pl}$			& X 		& 0.11	& 0.55 \\
					& X + R		& 0.22  & 0.13 \\
$L_X/L_{bol}$ with $M_{pl}$		& X 		& 0.18  & 0.37 \\
					& X + R		& -0.02 & 0.88 \\
$L_X$ with $a_{pl}^{-1}\times M_{pl}$	& X 		& 0.21  & 0.25 \\
					& X + R		& 0.31  & 0.03 \\
$L_X/L_{bol}$ with $a_{pl}^{-1}\times M_{pl}$	& X 	& 0.33  & 0.08 \\
					& X + R		& 0.09  & 0.51 \\ \hline
\end{tabular}
\caption{Correlation of X-ray luminosity and $L_X/L_{bol}$ with planetary parameters; X/R: XMM-Newton/ROSAT detections.}
\label{ranks}
\end{table}
%%%%%%%%%%%%%%%%%%%%%%%%%%%%%%%%%%%

\subsection{Comparison with field stars}

\begin{figure}
\includegraphics[width=0.5\textwidth]{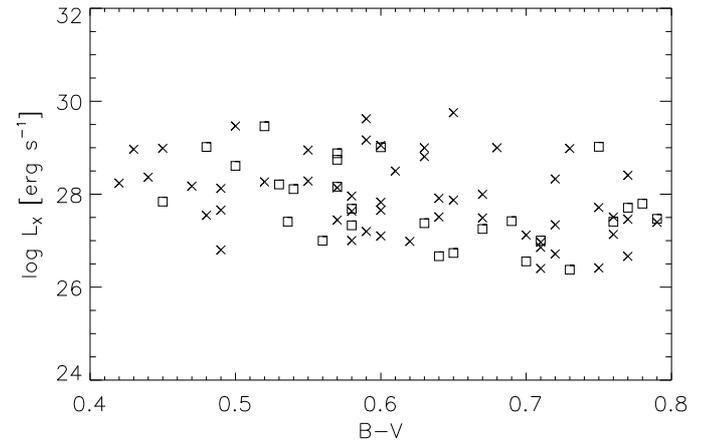}
\caption{X-ray luminosity of F and G type stars. Planet-bearing stars: squares, stars without planets: crosses.}
\label{fieldstars}
\end{figure}

To check for systematic differences, we compare our sample of planet-bearing stars with a sample of field stars of spectral type F and G as available from \cite{schmitt1997} from ROSAT observations. In Fig.~\ref{fieldstars} we show the X-ray luminosities of these stars over $B-V$ of both the planet-bearing and non-planet-bearing sample. A Kolmogorov-Smirnov test yields that the probability that both samples are drawn from the same parent distribution is 74\%. The values of the activity indicator $L_X/L_{bol}$ yield a probability of 23\% to be from the same distribution; this can be explained by the fact that stars of low activity are generally chosen for planet search programs.

%%%%%%%%%%%%%%%%%%%%%%%%%%%%%%%%%%%
%%%%%%%%%%%%%%%%%%%%%%%%%%%%%%%%%%%

\section{Star-Planet Interactions}\label{results}

Now we investigate our sample in detail for possible correlations of X-ray properties with planetary parameters.
%Detections und upper limits sind laut Kolmogorov-Smirnov für Lx und Lx/Lbol noch halbwegs vergleichbar (27% und 17%).

One expects that possible effects which giant planets might have on their host stars will strongly increase with decreasing orbital distance. Also, tidal as well as magnetic intercations should increase with the exoplanet's mass, assuming that larger exoplanets are capable of producing a stronger planetary magnetic field. Note that closer-in planets may rotate more slowly since they synchronize with their orbit, weakening their ability to generate magnetic fields \citep{griessmeier2004}; however, the details of planetary dynamos are not fully understood.

The most interesting quantity with regards to SPI in the X-ray regime is the activity indicator $L_X/L_{bol}$. The X-ray luminosity alone varies with stellar radius independently of the activity level, but $L_X/L_{bol}$ is independent of such radius-induced effects. Any planet-induced activity changes should therefore be evident in $L_X/L_{bol}$; a planet-induced variation in $L_X$ which would leave the ratio unchanged is rather unphysical, since $L_X/L_{bol}$ has typical values of $10^{-6}$ for our stars. A change in $L_{bol}$ would therefore need $10^6$ times more energy than the X-ray variation alone.

We study both the X-ray luminosity $L_X$ as well as the activity indicator $L_X/L_{bol}$ for correlations with the innermost planet's semimajor axis and mass. In Table~\ref{ranks} we give Spearman's $\rho$ rank correlation coefficient for various combinations. A value of $1$ ($-1$) means a perfect correlation (anticorrelation), $0$ means no correlation. The corresponding p-value gives the probability that the observed value of $\rho$ can be obtained by statistical fluctuations.

For the correlation analysis of $L_X/L_{bol}$, we exclude giants from our sample (HD~27442 and HD~62509), since they have very low $L_X/L_{bol}$ values due to their optical brightness. As well as for XMM-Newton detections alone as for XMM-Newton and ROSAT detection combined, we find no correlation of the semimajor axis with the stellar X-ray luminosity.

We find two possible correlations here: one of planetary mass with $L_X$ and a stronger one for $a_{pl}^{-1}\times M_{pl}$ with $L_X$. Stars with giant and close-in planets have higher X-ray luminosities than stars with small far-out planets. 
For $L_X/L_{bol}$. there is a correlation with $a_{pl}^{-1}\times M_{pl}$ present in the sample of XMM-Newton detections, but not in the larger sample of ROSAT and XMM-Newton detections, pointing towards the possibility that this correlation might be a statistical fluctuation. The probable reason for strong correlations in $L_X$, but weaker or absent ones in $L_X/L_{bol}$ is that there is also a strong ($>2\sigma$) correlation between $M_{pl}$ and $L_{bol}$: Stars with larger $L_{bol}$ are more massive, and around massive stars, giant planets are detected much more easily compared to small ones. Both correlations of planetary mass with $L_X$ and also $L_{bol}$ seem to cancel out in $L_X/L_{bol}$.

Another significant correlation worth mentioning exists between the planetary mass and the spectral type of the host star: small planets are prone to be found around stars of later types. This is basically the same trend we see between $M_{pl}$ and $L_{bol}$, since small planets are more easily detected around low-mass and therefore late-type stars.

%%%%%%%%%%%%%%%%%%%%%%%%%%%%%%%%%%%
\begin{figure}
\includegraphics[width=0.5\textwidth]{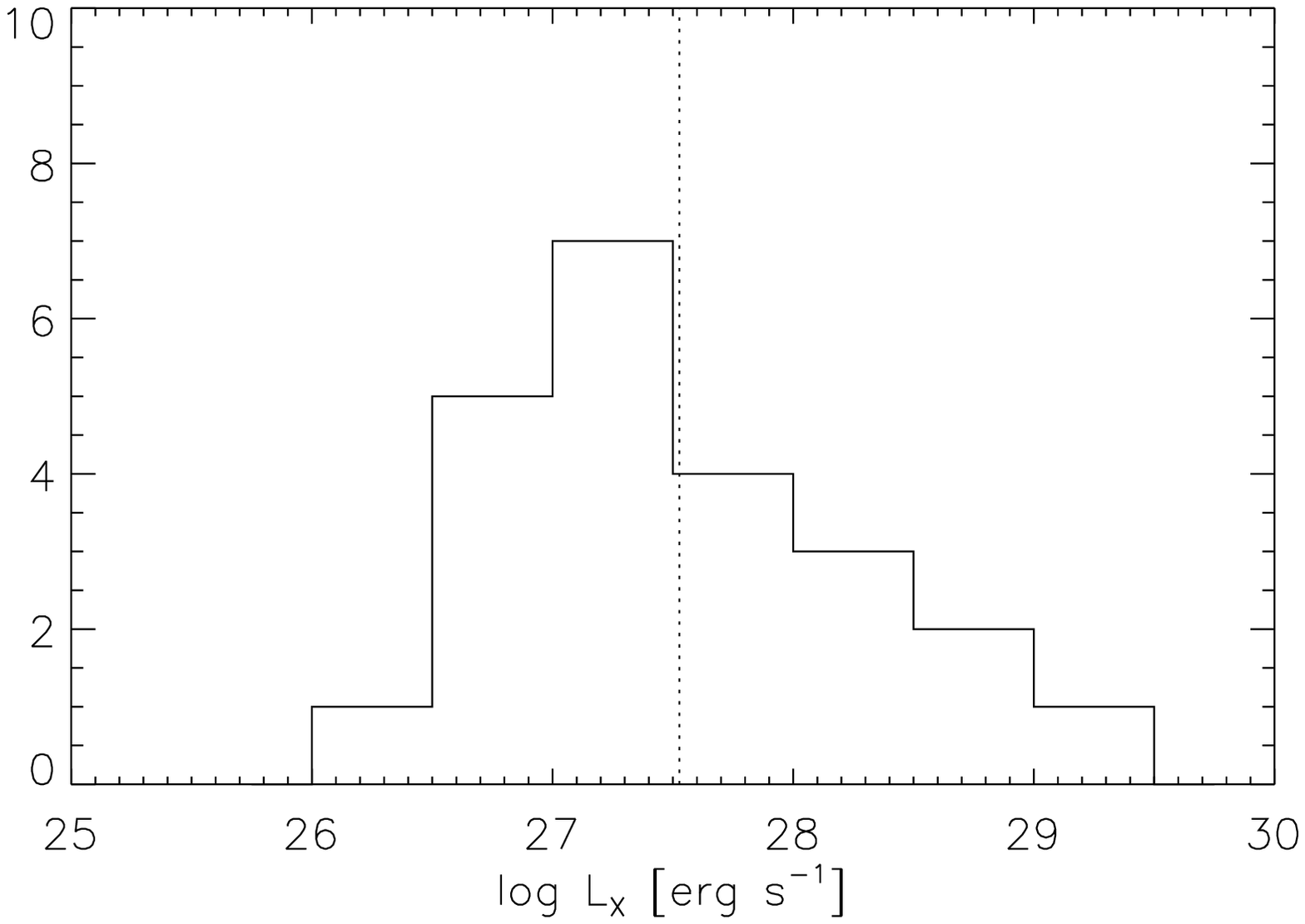}
\includegraphics[width=0.5\textwidth]{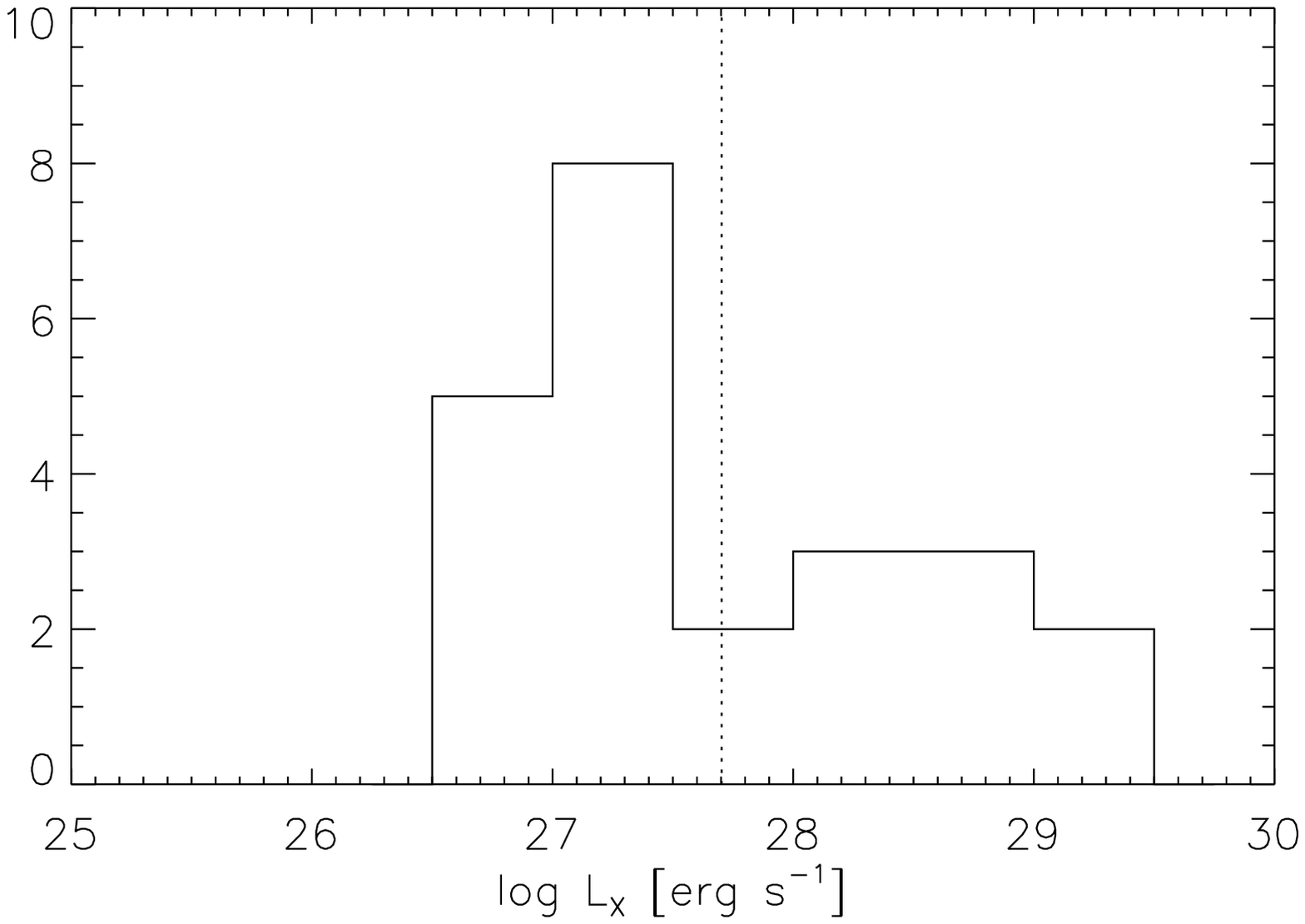}
\caption{Histograms of X-ray luminosities for X-ray detected stars within 30~pc with close-in ($a_{pl} \leq  0.2$~AU, upper panel) and far-out ($a_{pl} \geq 0.5$~AU, lower panel) planets. Mean $\log L_X$ values are indicated by dotted lines for both samples.}
\label{Histo_Lx}
\end{figure}
%%%%%%%%%%%%%%%%%%%%%%%%%%%%%%%%%%%

%%%%%%%%%%%%%%%%%%%%%%%%%%%%%%%%%%%
\begin{figure*}
\includegraphics[width=0.5\textwidth]{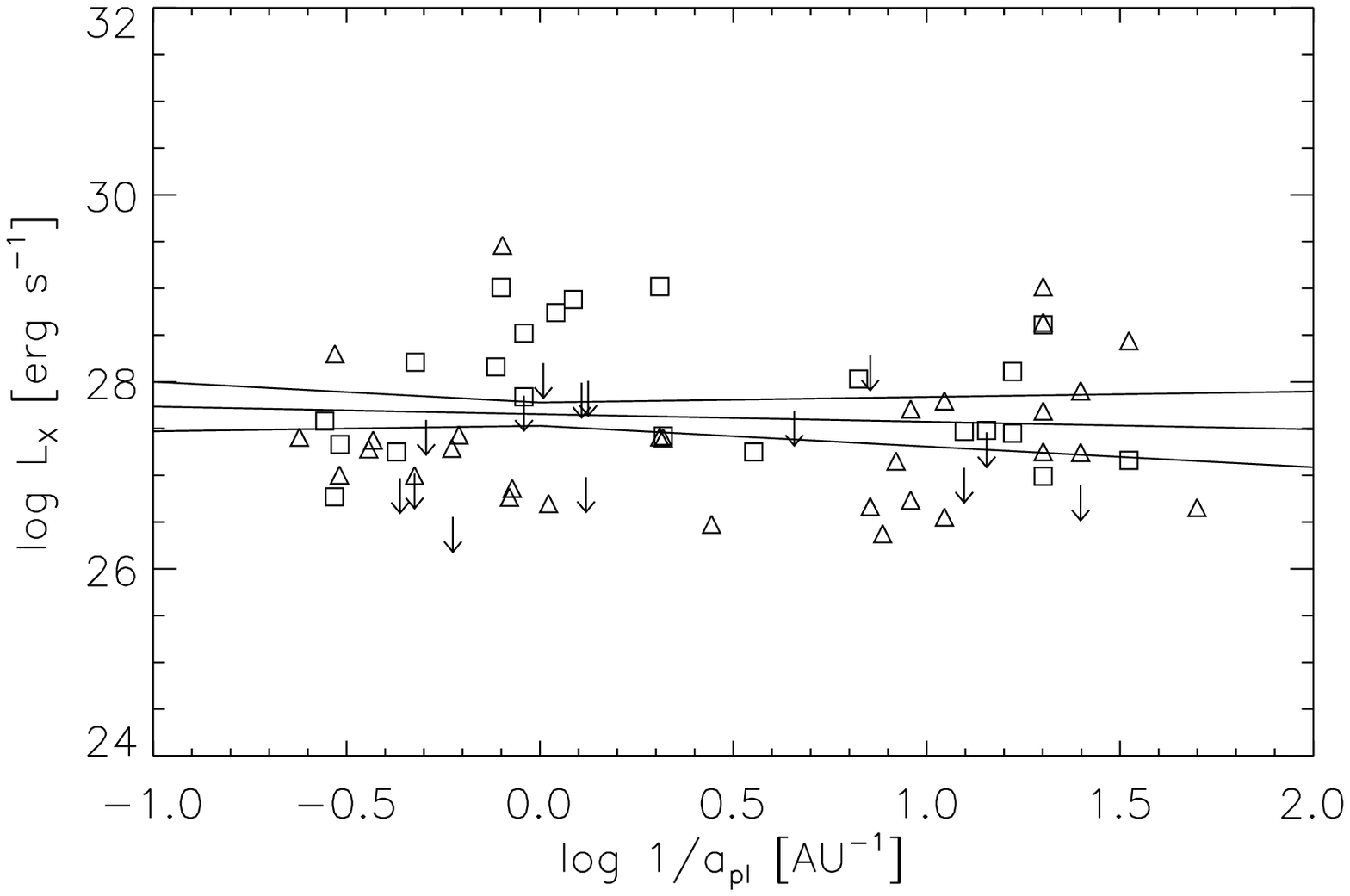}
\includegraphics[width=0.5\textwidth]{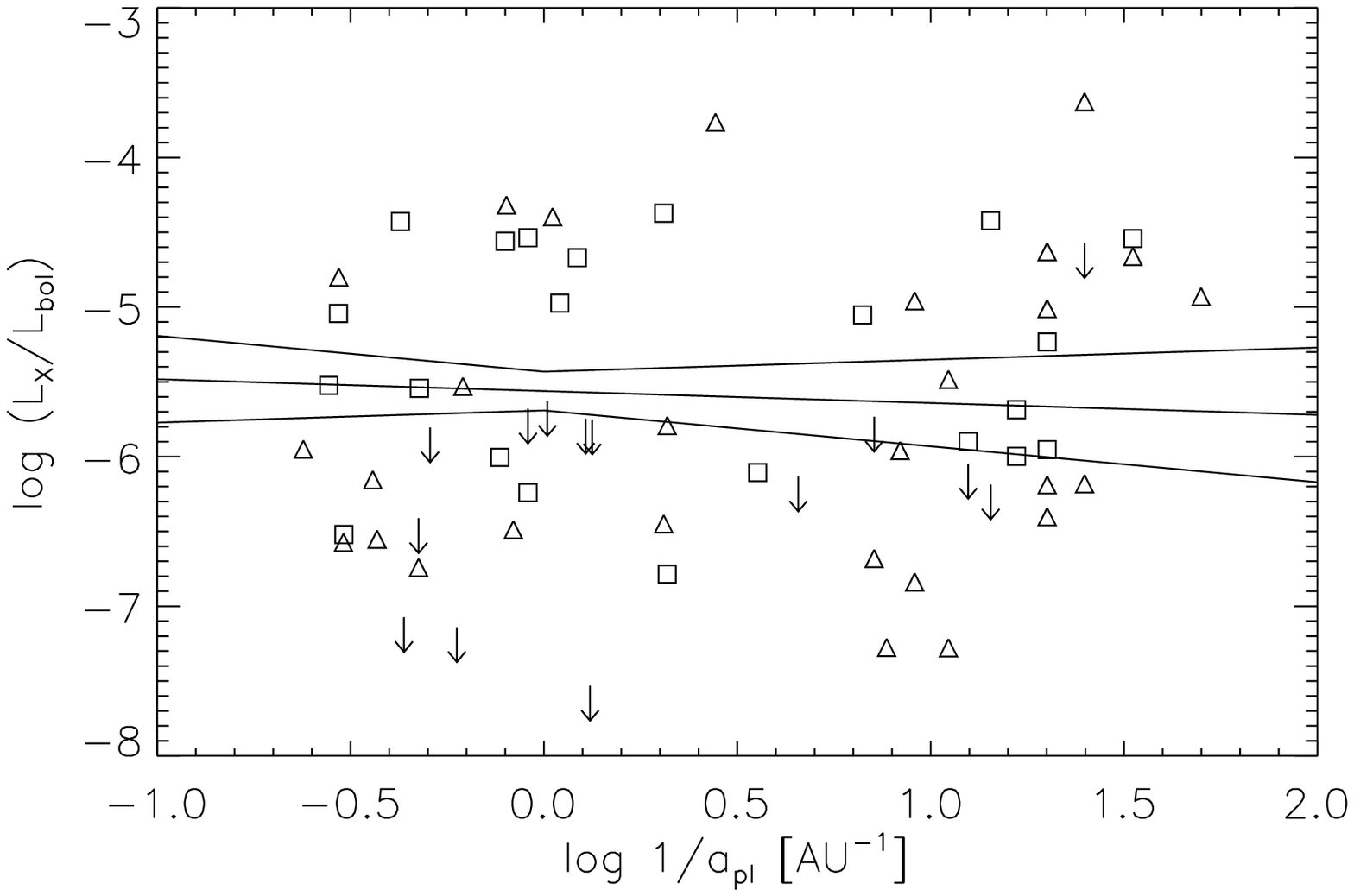}
\includegraphics[width=0.5\textwidth]{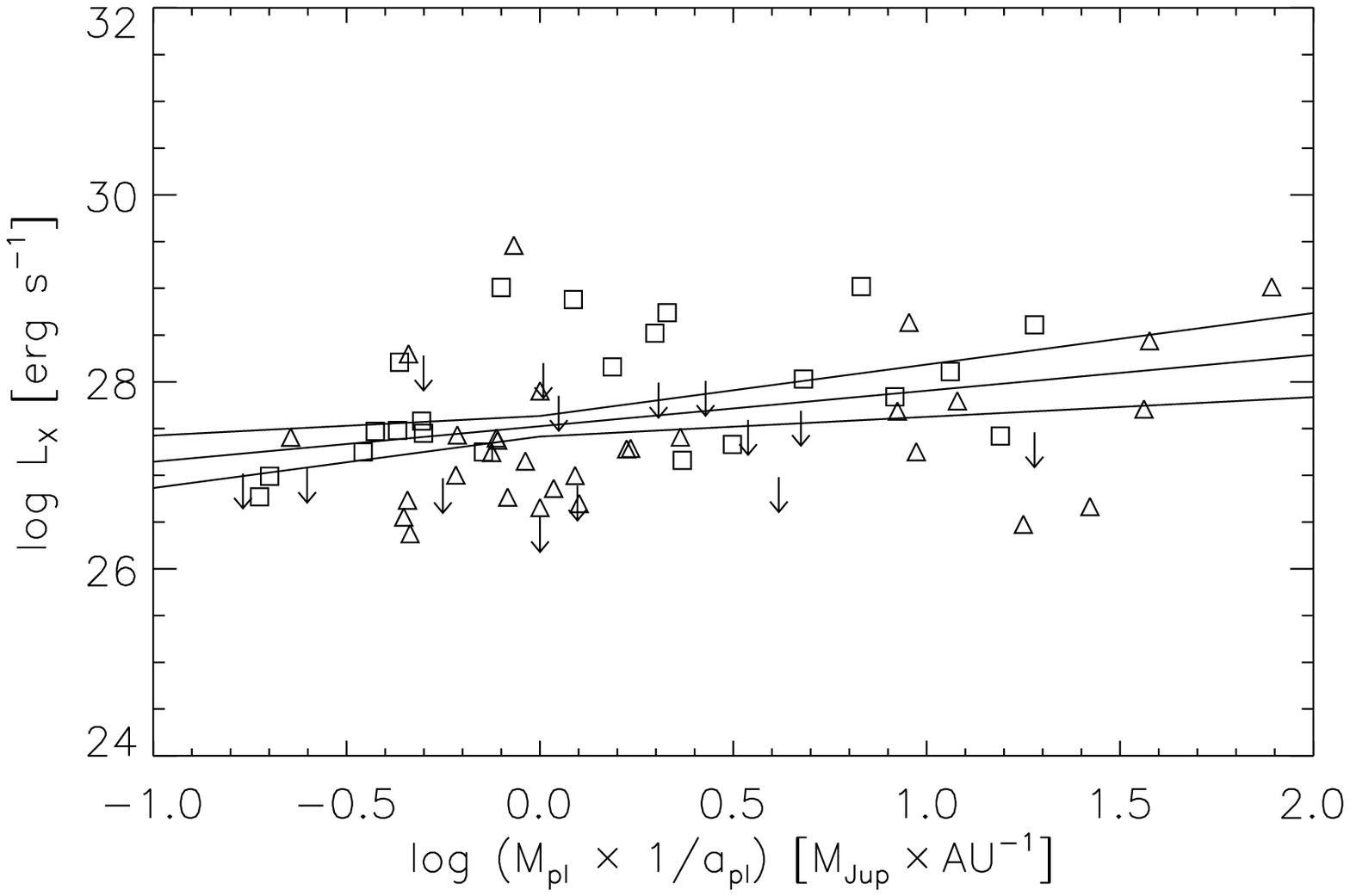}
\includegraphics[width=0.5\textwidth]{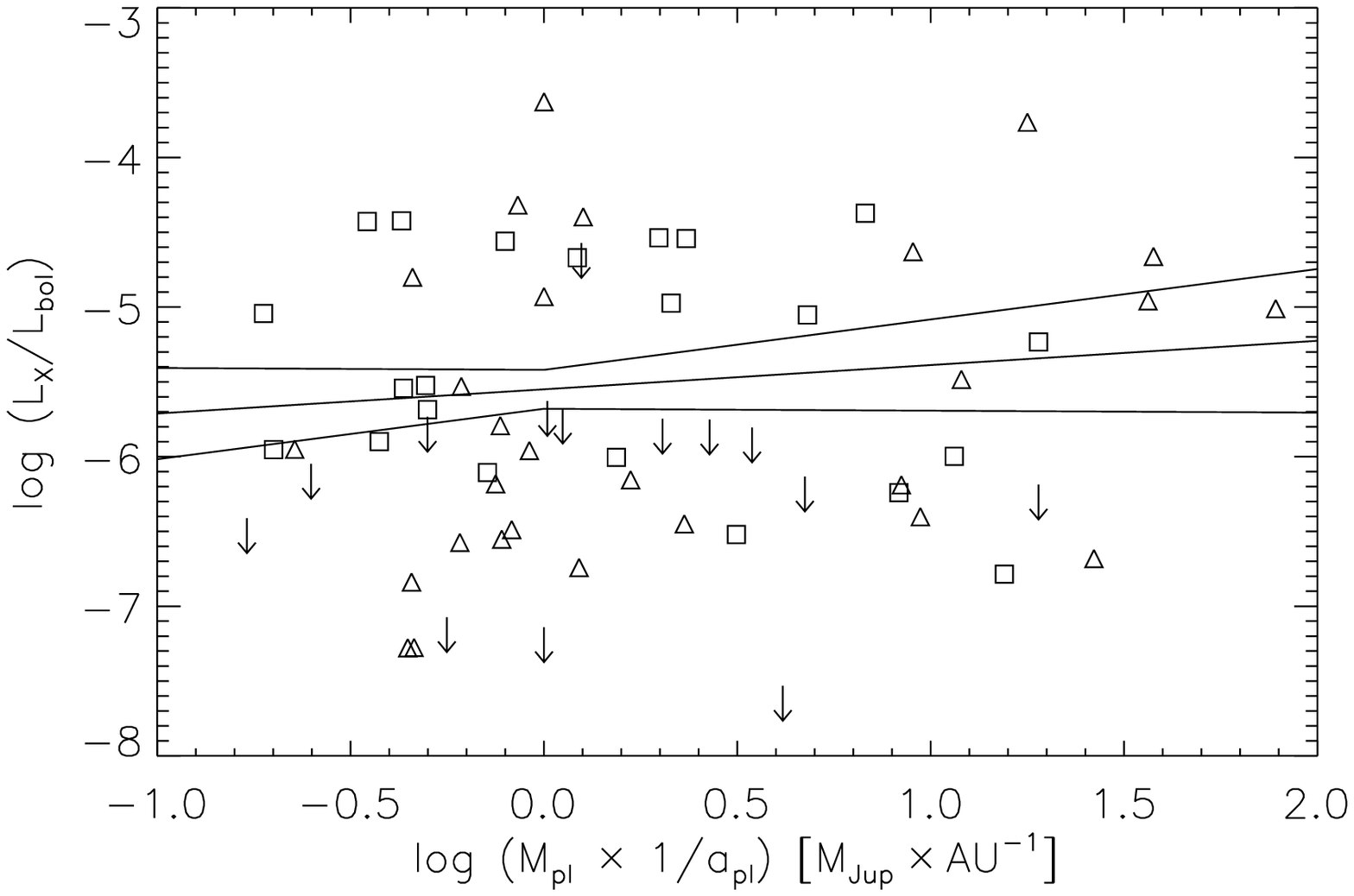}
\caption{X-ray luminosity and activity indicator $\log (L_X/L_{bol})$ as a function of $\log a_{pl}$ and $\log (a_{pl}^{-1}\times M_{pl})$, respectively. XMM-Newton detections: triangles, ROSAT detections: squares.}
\label{LxLbol}
\end{figure*}
%%%%%%%%%%%%%%%%%%%%%%%%%%%%%%%%%%%

To visualize these (non-)correlations, we perform linear regressions of $\log L_X$ and $\log (L_X/L_{bol})$ with either $\log a_{pl}$ or $\log (a_{pl}^{-1}\times M_{pl})$ by using the ''linmix\_err'' routine implemented in IDL. As already seen in the correlation analysis, we obtain slopes which are compatible with zero at $1\sigma$~level for all of the pictured cases except for $\log L_X$ over $\log (a_{pl}^{-1}\times M_{pl})$ (see Fig.~\ref{LxLbol}).

Independently of any linear trend, we can test if the $L_X$ values of stars with close-in and far-out planets stem from the same distribution. Fig.~\ref{Histo_Lx} shows the logarithmic X-ray luminosities of stars with planets within $a_{pl} \leq  0.2$~AU and stars with planets beyond $a_{pl} \geq 0.5$~AU. The means of both distributions are very similar and not distinguishable at $1\sigma$~level: $<\log L_{X\,close-in}>=27.52\pm0.72$~erg\,s$^{-1}$ and $<\log L_{X\,far-out}>=27.70\pm0.80$~erg\,s$^{-1}$. A Kolmogorov-Smirnov test yields a probability of 84\% for both samples being from the same distribution. However, the comparison of stars with {\em close-in, heavy} planets compared to {\em far-out light} ones yields that the probability for both samples to have the same parent distribution is very small with $<1\%$; the average X-ray luminosity is higher for stars with close-in, heavy planets with $<\log L_{X\,close-in\,heavy}>=27.91\pm0.76$~erg\,s$^{-1}$ compared to $<\log L_{X\,far-out\, light}>=27.41\pm0.73$~erg\,s$^{-1}$, but the means are compatible within $1\sigma$ errors.

\section{Properties of individual targets}

In the following, we give a short overview on our newly observed stars and their spectral properties. The stellar, planetary and X-ray properties of all planet-bearing stars which were observed with XMM-Newton are listed in Table~\ref{counts}.

\subsection{Individual targets}

Between May 2008 and April 2009 we observed a total of 20 planet-bearing stars in X-rays. One of these stars, {\em SCR~1845}, turned out to harbor a brown dwarf and not a planet; the X-ray characteristics of {\em SCR~1845} are discussed in a separate publication \citep{robradepoppenhaeger2010}. The X-ray properties of the remaining 19 stars are described briefly now.

\begin{figure}
\includegraphics[width=0.35\textwidth, angle=270]{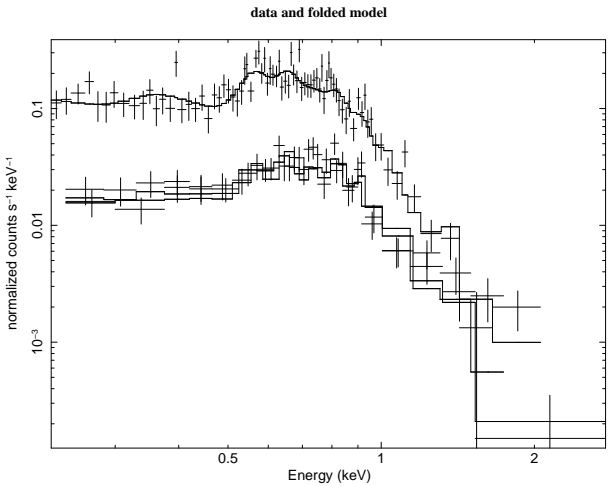}
\caption{Typical EPIC spectra (upper: PN, lower: MOS) of a planet-bearing star; here shown: {\em GL~86}.}
\label{EPICspectrum}
\end{figure}

{\em GJ~674}, {\em GL~86}, {\em GL~876}, {\em HD~102195}, {\em GJ~317}, {\em 55~Cnc} and {\em HD~99492} yielded sufficient signal-to-noise ratio for spectral fitting of the obtained EPIC spectra. They are all characterized by coronae with cool to moderate temperatures (details listed in section~\ref{spectralprop}). {\em GJ~674} shows one large and several smaller flares on timescales of ca.~5~ks. Also {\em GL~876} shows several short flares. The other stars show some variability around 15-25\% level. As an example, we show the EPIC spectrum and the corresponding two-temperature fit of {\em GL~86} in Figure~\ref{EPICspectrum}.

%%%%%%%%%%%%%%%%%%%%%%%%%%%%%%%%%%%
%%%%%%%%%%%%%%%%%%%%%%%%%%%%%%%%%%%

{\em HD~154345}, {\em HD~160691}, {\em HD~4308}, {\em HD~52265}, {\em HD~93083}, {\em 51~Peg}, {\em HD~27442}, {\em HD~114386} and {\em HD~114783} were detected in our exposures, but did not yield enough photons for spectral fitting. Where meaningful hardness ratios could be calculated from the numbers of counts, the stars proved to be soft X-ray sources, as one expects for nearby stars with low X-ray luminosities. Two of these targets show interesting characteristics: 

{\em HD~99492} is part of a binary system, consisting of a K2 dwarf, which is the planet-bearing star {\em HD~99492}, and a K0 subgiant, {\em HD~99491}. Previously, {\em HD~99492} was assigned an X-ray luminosity of $27.56$~erg\,s$^{-1}$ \citep{kashyapdrakesaar2008}, since the double system was unresolved in the corresponding ROSAT pointing. Our XMM-Newton pointing shows that {\em HD~99492} is actually the X-ray fainter part of the pair with an X-ray luminosity of only $26.93$~erg\,s$^{-1}$. 

{\em 51 Peg} is, despite a moderate X-ray luminosity of $26.8$~erg\,s$^{-1}$ which it exhibited in a ROSAT pointing from 1992, barely detected in a deep XMM-Newton observation. The XMM-Newton source photons are extremely soft, which explains its better visibility in ROSAT and additional recent Chandra data, since these instruments have a larger effective area at very low X-ray energies. Detailed analysis \citep{poppenhaeger2009} showed that the star is possibly in a Maunder minimum state. 

The stars {\em 16~Cyg~B}, {\em HD~111232}, {\em HD~217107} and {\em HD~164922} could not be detected in X-rays in our exposures. The upper limits for these stars were calculated for a confidence level of 99\%, following the lines of \cite{ayres2004} and point also towards low activity levels for these targets.

\subsection{Spectral properties}\label{spectralprop}

Seven of the stars we observed yielded sufficient signal to noise ratio for spectral fitting of their XMM-Newton EPIC data. The spectra of six stars can be adequately described by a thermal plasma with two temperature components and solar abundances; the spectral fitting was performed with Xspec~v12.5 and {\it apec} models. The derived spectral properties are listed in Table~\ref{apec}. They are mostly dominated by very cool plasma ($T\approx1$~MK) with small contributions from hotter plasma, only {\em HD~102195} and {\em GL~86} have stronger contributions from a hotter component around 4--5~MK.

%%%%%%%%%%%%%%%%%%%%%%%%%%%%%%%%%%%
\begin{table*}[ht!]
\begin{tabular}{l c c c c c c}
    \hline \hline
Parameter 		& GL 86 	& GL 876 	& HD 102195 	& GJ 317 	& 55 Cnc 	& HD 99492 \\
    \hline 
$T_1$ (keV)		& 0.11 		& 0.07 		& 0.09 		& 0.11		& 0.09		& 0.09		\\
$EM_1$ 			& 1.63 		& 0.42 		& 8.54		& 0.19		& 1.39		& 1.02		\\
$T_2$ (keV)		& 0.33 		& 0.39 		& 0.45		& 0.50		& 0.48		& 0.39		\\
$EM_2$ 			& 1.27 		& 0.06 		& 8.07		& 0.06		& 0.19		& 0.20		\vspace{0.2cm} \\ 
$\chi^2_{red}$ (d.o.f.)	& 1.02 (120)	& 1.44 (85) 	& 0.95 (152)	& 0.77 (22)	& 1.2 (11)	& 0.99 (8)	\\
$\log L_X$ (0.2-2.0~keV)& 27.6 		& 26.4		& 28.3		& 26.5		& 27.1		& 27.0		\\ \hline
\end{tabular}
\caption{Spectral modeling results derived from EPIC data; emission measure given in units of $10^{50}$~cm$^{-3}$.}
\label{apec}
\end{table*}
%%%%%%%%%%%%%%%%%%%%%%%%%%%%%%%%%%%

The spectrum of {\em GJ~674} cannot be fitted satisfactorily when assuming solar abundances, therefore we fit its EPIC spectra with two-temperature {\it vapec} models. {\em GJ~674} exhibits a flare during our observation, so we conducted the spectral analysis for the time interval of the flare as well as for quiescent times. The results are given in Table~\ref{gj674}. The temperature of both components rises considerably during the flare, as well as the total emission measure. Coronal abundances of the given elements are below solar values, consistent with the low photospherical iron abundance $[Fe/H]=-0.28$.

The X-ray luminosities derived from spectral modelling fit the ECF-derived results in Table~\ref{counts} well, justifying our error estimate of $\approx 30\%$, which we assumed in addition to statistical errors.

%%%%%%%%%%%%%%%%%%%%%%%%%%%%%%%%%%%
\begin{table}
\begin{tabular}{l c c }
    \hline \hline
Parameter 		& GJ 674 quiescence 	& GJ 674 flare \\ 
    \hline 
$T_1$ (keV)		& 0.14 			& 0.32 	\\
$EM_1$			& 0.71 			& 3.57 	\\	
$T_2$ (keV)		& 0.40 			& 0.81 	\\
$EM_2$ 			& 2.52	 		& 1.61 	\\
O			& 0.41 $\pm$ 0.07	& 0.38 $\pm$ 0.05\\
Ne			& 0.66 $\pm$ 0.13	& 0.38 $\pm$ 0.18\\
Mg			& 0.34 $\pm$ 0.14	& 0.51 $\pm$ 0.25 \\
Fe			& 0.29 $\pm$ 0.05	& 0.37 $\pm$ 0.12 \vspace{0.2cm}  \\ 
$\chi^2_{red}$ (d.o.f.)	& 1.12 (204)		& 1.00 (197) 	\\
$\log L_X$ (0.2-2.0~keV)& 27.5 		& 27.7		\\ \hline
\end{tabular}
\caption{Spectral modeling results for {\em GJ~674} derived from EPIC data; emission measure given in units of $10^{50}$~cm$^{-3}$.}
\label{gj674}
\end{table}
%%%%%%%%%%%%%%%%%%%%%%%%%%%%%%%%%%%

\section{Discussion}

\subsection{Interaction or selection?}\label{intsel}

In our data, we do not see any significant trend of the activity indicator $L_X/L_{bol}$ with the planetary semimajor axis, mass or a combination of both, in contrast to recent studies \citep{kashyapdrakesaar2008}. The only significant trend, as shown in section \ref{results}, is seen in the X-ray luminosity which is higher for stars with heavy close-in planets. Trying to explain this trend in $L_X$ without an accompanying trend in $L_X/L_{bol}$ by SPI is problematic. The sample stars have typical $L_X/L_{bol}$ values of $10^{-6}$. If planets induced higher $L_X$ levels, but constant $L_X/L_{bol}$ ratios, the amount of energy introduced by SPI would have to be $10^6$ times higher in $L_{bol}$ than in $L_X$. The $L_X$ excess of stars with close-in heavy planets compared to stars with far-out light ones is of the order of $5\times 10^{27}$~erg\,s$^{-1}$. Thus, the energy excess in $L_{bol}$ would have to be $\sim 10^{33}$~erg\,s$^{-1}$. Comparing this to the typical orbital energy of a Hot Jupiter ($\sim 10^{44}$~erg), this would lead to obviously unpyhsical timescales for the planet's orbital decay of only several thousand years.

However, there is also the possibility that the trend in $L_X$ is caused by selection effects: All but three of the planets in our sample have been detected with the radial velocity (RV) method. Stellar activity can mask the RV signal. Since the RV signal is strongest for close-in, heavy planets, we have a selection effect which favors detection of such planets around active stars. The key question is, do the data show an {\em additional} trend of $L_X$ with $a_{pl}^{-1}$ and $M_{pl}$ which is not induced by the selection effect and could be attributed to SPI?

\begin{figure}
\includegraphics[width=0.5\textwidth]{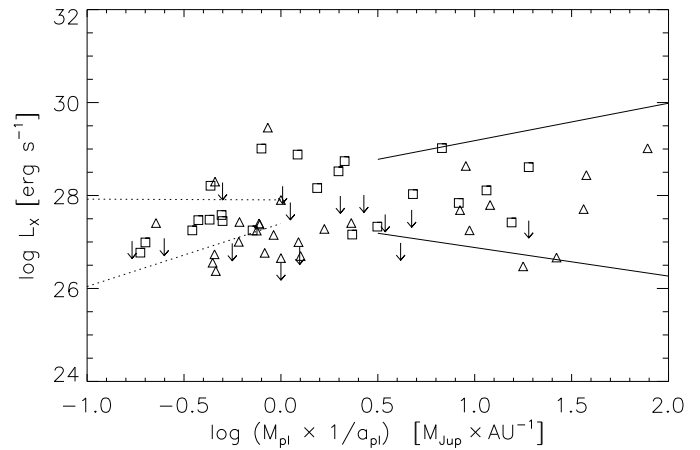}
\caption{Linear regression of $\log L_X$ over $\log (a_{pl}^{-1}\times M_{pl})$ for close-in heavy planets (solid) and far-out light planets (dotted). Both regressions overlap at $1\sigma$ level.}
\label{selection}
\end{figure}

To investigate this, we conduct two separate regression analyses on $\log L_X$ over $\log (a_{pl}^{-1}\times M_{pl})$ for close-in heavy planets and far-out light planets, respectively. We define close-in heavy planets by $\log (a_{pl}^{-1}\times M_{pl}) > 0.5$ (corresponding to a Jupiter-like planet at a maximum orbital distance of $\approx 0.3$~AU, for example) and far-out light planets by $\log (a_{pl}^{-1}\times M_{pl}) < 0$ (corresponding to a Jupiter-like planet at $1$~AU or a Saturn-like planet at $0.3$~AU). As shown in Fig.~\ref{selection}, both regressions overlap well at $1\sigma$ level, indicating that there is no additional activity enhancement effect measurable in this sample for close-in heavy planets other than the selection trend which also manifests itself in the subsample with far-out light planets.

%%%%%%%%%%%%%%%%%%%%%%%%%%%%%%%%%%%
%%%%%%%%%%%%%%%%%%%%%%%%%%%%%%%%%%%

%%%%%%%%%%%%%%%%%%%%%%%%%%%%%%%%%%%
            \begin{table*}[ht]
\setlength{\tabcolsep}{5pt}

               \tiny
      \begin{tabular}{l l r r r r r l r r r r r r}
    \hline
    \hline
       Star & Type & Dist. & $m_V$ & $B-V$ & $P_*$ & [Fe/H] & $a_{pl}$ & $M_{pl}$ & GTI &  net countrate$^a$ & $F_X$ & $\log L_X$ & $\log \frac{L_X}{L_{bol}}$\\
                                                                           & & (pc) & & & (d) & & (AU) & (M$_{J}$) & (s) &  (cts/ks) & (erg\,s$^{-1}$\,cm$^{-2}$) & 0.2-2~keV & \\

    \hline
                                   $\epsilon$ Eri & K2.0 V      &  3.2 &   3.73 & 0.88 &   11.2 & -0.10 &  3.39 &  1.55 & 10385 &  5611.5$\pm$   23.3$^b$ &   1.38E-11 & 28.22$\pm$ 0.12 & -4.88\\
                                                GJ 674 & M2.5   &  4.5 &   9.38 & 1.53 &   35.0 & -0.28 &  0.04 &  0.04 & 15183 &  1102.6$\pm$    8.6 &   2.16E-12 & 27.73$\pm$ 0.12 & -3.80\\
                                           GL 876 & M4.0 V      &  4.7 &  10.17 & 1.67 &   41.0 & -0.12 &  0.02 &  0.02 & 23436 &    48.9$\pm$    1.5 &   1.13E-13 & 26.48$\pm$ 0.13 & -5.11\\
                                                 VB 10 & M8.0 V      &  5.8 &   9.91 & 0.00 & - &  0.00 &  0.36 &  6.40 & 10810 &    23.2$\pm$    1.5 &   4.91E-14 & 26.30$\pm$ 0.14 & -1.79\\
                                                     GJ 317 & M3.5   &  9.2 &  12.00 & 1.53 & - & -0.23 &  0.95 &  1.20 & 11245 &    14.3$\pm$    1.5 &   3.30E-14 & 26.52$\pm$ 0.14 & -4.57\\
                                         HD 62509 & G5.0 IV-V   & 10.3 &   1.15 & 1.00 &  130.0 &  0.19 &  1.69 &  2.90 & 28759 &    31.2$\pm$    1.1 &   1.01E-13 & 27.11$\pm$ 0.13 & -6.63\\
                                            GL 86 & K1.0 V      & 11.0 &   7.40 & 0.77 &   31.0 & -0.24 &  0.11 &  4.01 & 13071 &   116.8$\pm$    3.1 &   2.94E-13 & 27.63$\pm$ 0.12 & -5.04\\
                                           55 Cnc & G8.0 V      & 13.0 &   5.95 & 0.87 &   42.7 &  0.29 &  0.04 &  0.03 &  8505 &    18.6$\pm$    1.6 &   5.77E-14 & 27.07$\pm$ 0.14 & -6.36\\
                                           47 UMa & G0.0 V      & 14.0 &   5.10 & 0.56 &   74.0 &  0.00 &  2.11 &  2.60 &  6196 &     2.6$\pm$    0.7 &   1.07E-14 & 26.40$\pm$ 0.21 & -7.34\\
                                           51 Peg & G5.0 V      & 14.7 &   5.49 & 0.67 &   37.0 &  0.20 &  0.05 &  0.47 & 25299 &     0.4$\pm$    0.2 &   1.70E-15 & 26.65$\pm$ 0.18 & -7.00\\
                                       $\tau$ Boo & M8.0 V      & 15.0 &   4.50 & 0.48 &    3.3 &  0.28 &  0.05 &  3.90 & 38251 &  1252.1$\pm$    5.7 &   3.21E-12 & 28.94$\pm$ 0.12 & -6.14\\
                                             HD 160691 & G3.0 IV-V   & 15.3 &   5.15 & 0.70 & - &  0.28 &  0.09 &  0.04 &  7046 &     3.6$\pm$    1.2 &   1.06E-14 & 26.47$\pm$ 0.16 & -7.36\\
                                             HD 190360 & G6.0 IV     & 15.9 &   5.71 & 0.73 & - &  0.24 &  0.13 &  0.06 &  2888 &     2.2$\pm$    1.4 &   5.23E-15 & 26.20$\pm$ 0.21 & -7.45\\
                                         HD 99492 & F7.0 V      & 18.0 &   7.57 & 1.01 &   45.0 &  0.36 &  0.12 &  0.11 & 19928 &     7.1$\pm$    0.6 &   2.44E-14 & 26.98$\pm$ 0.15 & -5.97\\
                                                14 Her & K0.0 V      & 18.1 &   6.67 & 0.90 & - &  0.43 &  2.77 &  4.64 &  5532 &    14.6$\pm$    2.9 &   3.25E-14 & 27.11$\pm$ 0.14 & -6.33\\
                                             HD 154345 & G8.0 V      & 18.1 &   6.74 & 0.76 & - & -0.11 &  4.19 &  0.95 &  3845 &    18.6$\pm$    2.4 &   5.46E-14 & 27.33$\pm$ 0.16 & -6.03\\
                                              HD 27442 & K2.0 III    & 18.1 &   4.44 & 1.08 & - &  0.20 &  1.18 &  1.28 &  4636 &     3.7$\pm$    1.3 &   1.23E-14 & 26.68$\pm$ 0.18 & -7.72\\
                                      $\beta$ Pic & A6.0 V      & 19.3 &   3.86 & 0.17 &    0.7 &  0.00 &  8.00 &  8.00 & 54896 &     0.2$\pm$    0.1$^c$ &   6.00E-16 & 25.40$\pm$ 0.15 & -9.09\\
                                             HD 189733 & K1.5   & 19.3 &   7.67 & 0.93 &   13.4 & -0.03 &  0.03 &  1.13 & 36271 &   110.3$\pm$    1.8 &   4.11E-13 & 28.26$\pm$ 0.12 & -4.84\\
                                              HD 217107 & G8.0 IV     & 19.7 &   6.18 & 0.72 &   37.0 &  0.37 &  0.07 &  1.33 &  5576 & $<$     6.0 & $<$   1.55E-14 & $<$ 26.86 & $<$ -6.79\\
                                        HD 195019 & G3.0 IV-V   & 20.0 &   6.91 & 0.64 &   22.0 &  0.08 &  0.14 &  3.70 &  8333 &     2.8$\pm$    0.8 &   6.44E-15 & 26.49$\pm$ 0.17 & -6.86\\
                                               16 Cyg B & G2.5 V      & 21.4 &   6.20 & 0.66 &   31.0 &  0.08 &  1.68 &  1.68 & 10768 & $<$     1.6 & $<$   5.42E-15 & $<$ 26.47 & $<$ -7.22\\
                                                   HD 164922 & K0.0 V      & 21.9 &   7.01 & 0.80 & - &  0.17 &  2.11 &  0.36 &  6955 & $<$     3.5 & $<$   1.21E-14 & $<$ 26.84 & $<$ -6.59\\
                                               HD 4308 & G0.0 V      & 21.9 &   6.54 & 0.65 & - & -0.31 &  0.11 &  0.05 &  7837 &     2.1$\pm$    0.7 &   7.89E-15 & 26.66$\pm$ 0.19 & -6.02\\
                                                  HD 114783 & K0.0   & 22.0 &   7.57 & 0.93 & - &  0.33 &  1.20 &  0.99 &  3583 &     2.1$\pm$    1.5 &   6.72E-15 & 26.59$\pm$ 0.19 & -6.66\\
                                             HD 216437 & G4.0 IV-V   & 26.5 &   6.06 & 0.63 & - &  0.00 &  2.70 &  2.10 &  3329 &     8.2$\pm$    1.9 &   1.89E-14 & 27.20$\pm$ 0.18 & -6.73\\
                                                   HD 20367 & G0.0   & 27.0 &   6.41 & 0.52 & - &  0.10 &  1.25 &  1.07 &  8861 &  1404.8$\pm$   12.6 &   2.76E-12 & 29.38$\pm$ 0.12 & -4.40\\
                                             HD 114386 & K3.0 V      & 28.0 &   8.80 & 0.90 & - &  0.00 &  1.62 &  0.99 &  3601 &     2.7$\pm$    1.2 &   7.19E-15 & 26.83$\pm$ 0.21 & -6.13\\
                                              HD 52265 & K0.0 III    & 28.0 &   6.30 & 0.54 & - &  0.11 &  0.49 &  1.13 &  6954 &     5.6$\pm$    1.0 &   1.82E-14 & 27.23$\pm$ 0.17 & -6.92\\
                                         HD 75289 & K3.0 V      & 28.9 &   6.35 & 0.58 &   16.0 &  0.29 &  0.05 &  0.42 &  6681 &     3.0$\pm$    0.7 &   1.21E-14 & 27.09$\pm$ 0.20 & -6.75\\
                                         HD 93083 & K2.0 V      & 28.9 &   8.33 & 0.94 &   48.0 &  0.15 &  0.48 &  0.37 &  7789 &     7.4$\pm$    1.3 &   1.67E-14 & 27.22$\pm$ 0.16 & -6.79\\
                                        HD 102195 & K0.0 V      & 29.0 &   8.06 & 0.83 &   12.0 &  0.05 &  0.05 &  0.45 & 13043 &   145.9$\pm$    3.4 &   2.87E-13 & 28.46$\pm$ 0.12 & -4.81\\
                                              HD 111232 & G8.0 V      & 29.0 &   7.61 & 0.68 &   30.7 & -0.36 &  1.97 &  6.80 &  6996 & $<$     3.2 & $<$   9.72E-15 & $<$ 26.99 & $<$ -6.41\\
                                              HD 70642 & G0.0 V      & 29.0 &   7.18 & 0.71 & - &  0.16 &  3.30 &  2.00 & 10935 &     3.0$\pm$    0.7 &   6.68E-15 & 26.83$\pm$ 0.17 & -8.08\\
                                        HD 130322 & K0.0 V      & 30.0 &   8.05 & 0.78 &    8.7 & -0.02 &  0.09 &  1.08 &  4194 &    16.7$\pm$    2.2 &   3.87E-14 & 27.62$\pm$ 0.16 & -5.66\\
              \hline
       \end{tabular}
\caption{Stellar and planetary parameters of planet-bearing stars within 30 pc, as observed by XMM-Newton. Stellar and planetary parameters taken from www.exoplanet.eu, bolometric luminosities calculated from $m_V$ with bolometric corrections from \cite{flower1996}.
\newline $^a$ PN, 0.2-2~keV  
\newline $^b$ MOS1 countrate given, since PN detector suffered from pile-up for this observation 
\newline $^c$ Data taken from \cite{HempelRobrade2005}, combined MOS countrate given, since PN detector was optically contaminated }
      \label{counts}
        \end{table*}

%%%%%%%%%%%%%%%%%%%%%%%%%%%%%%%%%%%

%%%%%%%%%%%%%%%%%%%%%%%%%%%%%%%%%%%

            \begin{table*}[ht]

\begin{center}
\setlength{\tabcolsep}{9pt}

               \tiny
      \begin{tabular}{l l r r r r r l r r r r r r}
    \hline
    \hline
                                Star & Type & Dist. & $m_V$ & $B-V$ & $P_*$ & [Fe/H] & $a_{pl}$ & $M_{pl}$ & $\log L_X$ & $\log \frac{L_X}{L_{bol}}$\\
                                                                                                               & & (pc) & & & (d) & & (AU) & (M$_{J}$) &0.1-2.4~keV & \\
    \hline
                                                      GJ 832 & M1.5   &  4.9 &   8.67 & 1.46 & - & -0.31 &    3.40 &  0.64 & 26.77$\pm$ 0.21 & -9.09\\
                                                   GL 581 & M3.0   &  6.3 &  10.55 & 1.60 &   84.0 & -0.33 &    0.04 &  0.05 & $<$ 26.89 & $<$ -4.57\\
                                                Fomalhaut & A3.0 V      &  7.7 &   1.16 & 0.09 & - &  0.00 &  115.00 &  3.00 & $<$ 25.90 & $<$ -8.88\\
                                                      GJ 849 & M3.5   &  8.8 &  10.42 & 1.52 & - &  0.00 &    2.35 &  0.82 & 27.25$\pm$ 0.26 & -6.73\\
                                         HD 285968 & M2.5 V      &  9.4 &   9.97 & 1.51 &   38.9 & -0.10 &    0.07 &  0.03 & 27.48$\pm$ 0.28 & -6.13\\
                                                 GJ 436 & M2.5   & 10.2 &  10.68 & 1.52 &   45.0 & -0.32 &    0.03 &  0.07 & 27.16$\pm$ 0.34 & -6.36\\
                                                HD 3651 & K0.0 V      & 11.0 &   5.80 & 0.92 & - &  0.05 &    0.28 &  0.20 & 27.25$\pm$ 0.23 & -5.66\\
                                               HD 69830 & K0.0 V      & 12.6 &   5.95 & 0.79 & - & -0.05 &    0.08 &  0.03 & 27.47$\pm$ 0.30 & -4.84\\
                                               HD 40307 & K2.5 V      & 12.8 &   7.17 & 0.93 & - & -0.31 &    0.05 &  0.01 & 26.99$\pm$ 0.28 & -7.36\\
                                              HD 147513 & G3.0 V      & 12.9 &   5.37 & 0.60 & - & -0.03 &    1.26 &  1.00 & 29.01$\pm$ 0.16 & -3.80\\
                                    $\upsilon$ And & F8.0 V      & 13.5 &   4.09 & 0.54 &   12.0 &  0.09 &    0.06 &  0.69 & 28.11$\pm$ 0.22 & -6.86\\
                                           $\gamma$ Cep & K2.0 V      & 13.8 &   3.22 & 1.03 & - &  0.00 &    2.04 &  1.60 & 26.96$\pm$ 0.20 & -7.34\\
                                            HR 810 & G0.0 V      & 15.5 &   5.40 & 0.57 &    7.0 &  0.25 &    0.91 &  1.94 & 28.74$\pm$ 0.21 & -7.45\\
                                              HD 128311 & K0.0   & 16.6 &   7.51 & 0.99 &   11.5 &  0.08 &    1.10 &  2.18 & 28.52$\pm$ 0.21 & -4.57\\
                                                HD 7924 & K0.0 V      & 16.8 &   7.19 & 0.82 & - & -0.15 &    0.06 &  0.03 & 27.45$\pm$ 0.29 & -6.75\\
                                               HD 10647 & F8.0 V      & 17.3 &   5.52 & 0.53 & - & -0.03 &    2.10 &  0.91 & 28.21$\pm$ 0.17 & -4.88\\
                                          $\rho$ CrB & G0.0 V      & 17.4 &   5.40 & 0.61 &   19.0 & -0.24 &    0.22 &  1.04 & $<$ 27.69 & $<$ -6.13\\
                                                GJ 3021 & G6.0 V      & 17.6 &   6.59 & 0.75 & - &  0.20 &    0.49 &  3.32 & 29.02$\pm$ 0.21 & -7.00\\
                                               HD 87833 & K0.0 V      & 18.1 &   7.56 & 0.97 & - &  0.09 &    3.60 &  1.78 & 27.58$\pm$ 0.20 & -6.79\\
                                         HD 192263 & K2.0 V      & 19.9 &   7.79 & 0.94 &   27.0 & -0.20 &    0.15 &  0.72 & 28.03$\pm$ 0.35 & -4.81\\
                                               HD 39091 & G1.0 IV     & 20.5 &   5.67 & 0.58 & - &  0.09 &    3.29 & 10.35 & 27.33$\pm$ 0.20 & -6.03\\
                                                   HD 142 & G1.0 IV     & 20.6 &   5.70 & 0.52 & - &  0.04 &    0.98 &  1.00 & $<$ 28.20 & $<$ -5.63\\
                                               HD 33564 & F6.0 V      & 21.0 &   5.08 & 0.45 & - & -0.12 &    1.10 &  9.10 & 27.84$\pm$ 0.30 & -6.66\\
                                                HD 210277 & G0.0 V      & 21.3 &   6.63 & 0.71 & - &  0.19 &    1.10 &  1.23 & $<$ 27.85 & $<$ -5.68\\
                                            70 Vir & G4.0 V      & 22.0 &   5.00 & 0.69 &   31.0 & -0.03 &    0.48 &  7.44 & 27.42$\pm$ 0.28 & -6.33\\
                                               HD 19994 & F8.0 V      & 22.4 &   5.07 & 0.57 & - &  0.23 &    1.30 &  2.00 & 28.16$\pm$ 0.28 & -6.41\\
                                                HD 134987 & G5.0 V      & 25.0 &   6.45 & 0.70 & - &  0.23 &    0.78 &  1.58 & $<$ 27.99 & $<$ -5.75\\
                                                 HD 16417 & G1.0 V      & 25.5 &   5.78 & 0.67 & - &  0.19 &    0.14 &  0.07 & $<$ 28.28 & $<$ -5.73\\
                                                 HD 60532 & F6.0 IV-V   & 25.7 &   4.45 & 0.48 & - & -0.42 &    0.76 &  3.15 & $<$ 26.98 & $<$ -7.53\\
                                                HD 181433 & K3.0 IV     & 26.1 &   8.38 & 1.04 & - &  0.33 &    0.08 &  0.02 & $<$ 27.08 & $<$ -6.05\\
                                                 HD 30562 & F8.0 V      & 26.5 &   5.77 & 0.63 & - &  0.24 &    2.30 &  1.29 & $<$ 26.97 & $<$ -7.07\\
                                         HD 179949 & F8.0 V      & 27.0 &   6.25 & 0.50 &    9.0 &  0.22 &    0.05 &  0.95 & 28.61$\pm$ 0.25 & -4.40\\
                                                   HD 150706 & G0.0   & 27.2 &   7.03 & 0.57 & - & -0.13 &    0.82 &  1.00 & 28.88$\pm$ 0.19 & -5.04\\
                                                 HD 82943 & G0.0 V      & 27.5 &   6.54 & 0.62 & - &  0.27 &    0.75 &  2.01 & $<$ 28.01 & $<$ -5.75\\
              \hline
       \end{tabular}

\caption{Stellar and planetary parameters of planet-bearing stars within 30 pc, as observed by ROSAT. Stellar and planetary parameters taken from www.exoplanet.eu, bolometric luminosities calculated from $m_V$ with bolometric corrections from \cite{flower1996}. X-ray luminosities taken from \cite{kashyapdrakesaar2008}, except for upper limits for {\em HD~16417}, {\em HD~30562}, {\em HD~181433} and {\em HD~60532}, which were calculated from original data.}
\end{center}
 \label{rosatcounts}
        \end{table*}

%%%%%%%%%%%%%%%%%%%%%%%%%%%%%%%%%%%

\subsection{Is there evidence for {\em coronal} SPI?}

There are two different scenarios for SPI: tidal and magnetic interaction (see for example \cite{CuntzSaar2000}). {\em Tidal interaction} will affect motions in the stellar convection zone as well as the flow of plasma in the outer atmospheric layers. If stellar rotation and the planetary orbital motion are not synchronous, tidal bulges should rise and subside on the star, causing additional turbulence at the footpoints of magnetic loops, leading to higher flaring rates, or causing outer layers of the star to corotate with the planet, which might enhance the stellar dynamo if $P_{orb} > P_{*}$. {\em Magnetic interaction} is thought to be able to enhance the stellar activity in several ways: if planets are close enough to their host stars to be located inside the star's Alfv\'{e}n radius, a Jupiter-Io-like interaction can form where the planet is connected with the star via fluxtubes which heat the stellar atmosphere at their footpoints. Alternatively, magnetic reconnection events of the stellar and planetary magnetic field lines might supply additional energy to the stellar atmosphere. Also the mere presence of the planetary magnetic field itself might affect stellar wind formation and coronal densities, as a recent study \citep{CohenDrake2009} suggests.

Indications for SPI signatures in stellar {\em chromospheres} were found by \cite{Shkolnik2005} for two out of 13 stars, namely {\em HD~179949} and {\em $\upsilon$~And}, both stars with Hot Jupiters. The spectra of those stars showed periodic peaks in the \ion{Ca}{ii} H and K line emissions, common chromospheric activity indicators. The amplitude of the variation was 2.5\% for {\em HD~179949} and 0.7\% for {\em $\upsilon$~And} in the K line flux compared to a mean spectrum of the respective star and appeared once per orbital period of the planet in several years. However, the peaks changed to a once-per-stellar-rotation cycle in other years, suggesting an "On/Off"-behavior of chromospheric SPI. The fact that those peaks appeared only once per orbital period points towards magnetic SPI, since in a tidal SPI scenario one would expect two peaks in that time, which is not backed up by the Shkolnik data. 

A first statistical study on possible X-ray flux enhancements due to Hot Jupiters has been conducted by \cite{kashyapdrakesaar2008}. They claim to have found strong evidence that stars with Hot Jupiters are on average more X-ray active than stars with distant planets. Their study uses  main-sequence planet-bearing stars which were known at the time of writing, but the X-ray detection rate among these stars was only approximately one third, so that the authors had to include a large number of upper limits in their analysis and used Monte Carlo simulations on the X-ray luminosities of their sample. Their analysis suggests that stars with planets closer than $0.15$~AU have on average four times higher X-ray luminosity than stars with planets at distances larger than $1.5$~AU. They try to account for selection effects by regarding the trend of $L_X/L_{bol}$ with $a_{pl}$ as selection-induced and the remaining trend in $L_X$ with $a_{pl}$ as planet-induced, which leads to remaining $L_X$-ratio of stars with close-in and far-out planets of $\approx 2$, with an overlap of the (simulated) $L_X$-distributions at $1\sigma$ level. 

We do not see a significant difference of $L_X$-distributions in dependence on $a_{pl}$ in our sample as shown in Fig.~\ref{Histo_Lx}. There is also no significant trend of $L_X/L_{bol}$ with $a_{pl}$ evident in our data. We do not try to correct artificially for selection effects, since these are various and interdependent: since stellar activity masks the planet-induced radial-velocity signal, small far-out planets are more easily detected around very inactive stars. Similarly, those planets are also easier to detect around low-mass and therefore late-type stars, but very late-type stars have again higher $L_X/L_{bol}$ values than earlier-type and therefore heavier stars. A quantitative estimate of activity-related selection effects is therefore extremely difficult. But as shown in section~\ref{intsel}, there is no significant {\em additional} effect on activity visible in our data which can be attributed to the influence of massive close-in planets.

There has also been an effort to measure coronal SPI for an individual target: the star {\em HD~179949} by \cite{SaarSPI2008}, which showed the largest SPI signatures in chromospheric data so far. The star's X-ray flux was measured several times during May 2005; in September 2005, the star was in an "On"-state of SPI as seen in the chromosphere. The measured X-ray fluxes above 0.3~keV vary by $\pm 15\%$, a typical level also for intrinsic stellar X-ray variability. When folded with the orbital period, the profile of X-ray variability does not match the one seen in chromospheric data very well; interpretations of variability with the stellar rotation period are also possible. To see how chromospheric and coronal variability compare with each other for this star, we can do a rough estimate: a variation of $2.5\%$ in the \ion{Ca}{ii} K line compared to the mean stellar spectrum should translate into something of the same relative order of magnitude for the Mount Wilson S~index. The S~index variation should even be a bit smaller than $2.5\%$, since the S index averages over the H and K line, and the H line is less sensitive to activity effects. If we compare this to stars with known activity cycles such as {\em 61~Cyg}, one finds there \citep{HempelmannRobrade2006} that the S~index during one stellar activity cycle varies by $\pm 15\%$, while the X-ray flux in the $0.2-2.0$~keV band varies by $\pm 40\%$. A similar ratio of X-ray to \ion{Ca}{ii} fluxes yields for {\em HD~179949} an expected X-ray variation of $\approx 7\%$, less than a typical intrinsic variability level for late-type main-sequence stars. However, this ratio between X-ray and \ion{Ca}{ii} fluxes should only apply if activity enhancement via SPI works via similar mechanisms as normal stellar activity does, which is not necessarily the case given the possibility of Jupiter-Io-like interaction scenarios.

A basic quantitative scenario for coronal magnetic SPI has been proposed by \cite{CohenDrake2009}, suggesting that the presence of a close-in planetary magnetosphere hinders the expansion of the stellar magnetic field and the acceleration of the stellar wind, causing a higher plasma density in a coronal "hot spot". In their model, the hot spot leads to variations of the X-ray flux of $\approx 30\%$ when rotating in and out of view, and compared to a setting without a planet, the overall X-ray flux is enhanced by a factor of $\approx1.5$ for a stellar dipole field and a factor of $\approx15$ for a stellar magnetic field like the Sun's in an activity maximum. If we compare this to our sample, we can conclude that an activity enhancement of more than one order of magnitude is not a common effect in stars with Hot Jupiters, since such an activity overshoot for stars with Hot Jupiters would yield significantly different results in our sample's trend of $L_X$ with $\log (a_{pl}^{-1}\times M_{pl})$.

In summary, we can conclude from our analysis that there is no major average activity enhancement in the corona of stars which is induced by their planets. Any trends seen in our sample seem to be dominated by selection effects. 

\subsection{Promising individual targets}

SPI has been claimed for a handful of targets in chromospheric emissions at selected times. Even if there has been no stringent detection of corresponding coronal SPI yet for these stars, simultaneous observations in the optical and X-ray regime yield precious insight into the interplay of stellar and planetary magnetic fields. The most promising candidates for such coordinated searches are stars with close-in and massive planets which are rather X-ray bright to allow feasible observation plans. This identifies the stars {\em HD~102195}, {\em HD~130322}, {\em HD~189733}, {\em $\tau$~Boo}, {\em $\upsilon$~And}, {\em HD~179949} as promising targets. Four of these, namely {\em HD~189733}, {\em $\tau$~Boo}, {\em $\upsilon$~And} and {\em HD~179949}, have been investigated for chromospheric SPI before \citep{Shkolnik2005}. Out of these, {\em $\upsilon$~And} and {\em HD~179949} did show chromospheric activity enhancement in phase with the planetary orbit; the other two stars showed indications for increased {\em variability} of chromospheric activity with $P_{orb}$ \citep{WalkerCroll2008, Shkolnik2005}. The remaining two stars, {\em HD~102195} and {\em HD~130322}, have not been analyzed in detail for chromospheric SPI yet, but might as well be interesting targets for combined chromospheric and coronal SPI searches.

\section{Conclusions}
We analyzed a sample of all known planet-bearing stars in the solar neighborhood for X-ray activity and possible manifestations of coronal Star-Planet-Interactions (SPI) in dependence of the planetary parameters mass and semimajor axis. Our main results are summarized as follows:
\begin{enumerate}
\item In our sample of $72$ stars, there are no significant correlations of the activity indicator $L_X/L_{bol}$ with planetary mass or semimajor axis.
\item However, we do find a correlation of the X-ray luminosity with the product of planetary mass and reciprocal semimajor axis. Massive close-in planets are often found around X-ray brighter stars.
\item This dependence can be ascribed to selection effects: The radial velocity method for planet detections favors small and far-out planets to be detected around low-activity, X-ray dim stars. Additionally, if SPI induced an excess in $L_X$ without changing the $L_X/L_{bol}$ ratio, SPI would need to cause extremely high energy input in $L_{bol}$, leading to unrealistically short decay times for the planetary orbit.
\item There is no {\em additional} effect detectable in $L_X$ which could be attributed to coronal manifestations of SPI.
\item Coronal SPI might still be observable for some individual promising targets, preferably in coordinated observations of the targets' coronae and chromospheres.
\end{enumerate}

\begin{acknowledgements}
K.~P.~ and J.~R.~ acknowledge financial support from DLR grants 50OR0703 and 50QR0803. This work is based on observations obtained with {\em XMM-Newton} and makes use of the {\em ROSAT} Data Archive.
\end{acknowledgements}

% bibtex in der Konsole ausführen
\bibliographystyle{aa}
\bibliography{14245.bib}

\end{document}